\documentclass[runningheads]{svmult}
\usepackage{makeidx}   
\usepackage{graphicx}  
\usepackage{subeqnar}  
\usepackage{multicol}  
\usepackage{physprbb}  
\makeindex             

\def\mic{$\mu$m\ }
\def\etal{{et al.}\ }
\def\sles{\lower2pt\hbox{$\buildrel {\scriptstyle <}
   \over {\scriptstyle\sim}$}}
\def\teff{T$_{\rm eff}\,$}
\def\teffs{T$_{\rm eff}$s$\,$}
\def\apj{Astrophys.~J.\ }
\def\aj{Astron.~J.\ }
\def\araa{Annu.~Rev.~Astron.~Astrophys.\ }

\begin{document}
\title*{Alkali Metals and the Color of Brown Dwarfs}
\toctitle{Alkali Metals and the Color of Brown Dwarfs}
\titlerunning{Alkali Metals and the Color of Brown Dwarfs}
\author{Adam Burrows\inst{1}}
\authorrunning{Adam Burrows}
%
%
\institute{$^{1}$ Department of Astronomy and Steward Observatory, \\
The University of Arizona, Tucson, AZ 85721, USA}

\maketitle              

\begin{abstract}

I summarize some of the consequences for the optical and very-near-infrared spectra of T dwarfs (in
particular) and brown dwarfs (in general) of their possible dominance by the neutral alkali metal 
lines.  As a byproduct of this study, I estimate the true optical color of ``brown'' dwarfs.

\end{abstract}

\section{Introduction}

The early discovery phase for L dwarfs and T dwarfs has ended and a major
focus is now on their characterization.  The atmospheres of brown dwarfs are dominated
by H$_2$, H$_2$O, CH$_4$, NH$_3$, the neutral alkali metals, and grains, but how
theory translates this basic knowledge into effective temperatures, gravities, and
compositions has yet to be determined.  Establishing the spectral and color diagnostics
that are most appropriate for L/T studies is complicated by ambiguities in the cloud/grain
models and a paucity of opacity data.  In particular, though T dwarfs are being informally
defined by their methane features at 1.7 \mic and 2.2 \mic, the methane database itself
is far from complete.  The methane opacities on the red side of the $H$ band are certainly
in error by a factor of 3 to 5 (witness Gliese 229B\cite{leg99}) and the hot bands are completely missing.  The latter
means that even the sign of the opacity's dependence upon temperature can be in error.  
Nevertheless, there has been great overall progress towards understanding what makes 
these objects unique and what their spectra are telling us.  In this paper, I sidestep 
a comprehensive study of these issues and summarize three interesting topics in brown dwarf
theory that have emerged of late.  They are 1) what determines T dwarf spectra shortward of 1.0 micron,
2) what is the true color of a ``brown" dwarf, and 3) what is the effect of heavy element depletion (``rainout'')
on the abundance profiles of the neutral alkali metal atoms.  
A subtext of this contribution is the central importance of
the alkali metals in spectrum formation.

\section{The Short-Wavelength Spectra of T Dwarfs}

Employing the scheme of Burrows, Marley, and Sharp\cite{bms}
(hereafter BMS), we can derive the neutral alkali opacities as a function of wavelength.  Figure 1 depicts the
abundance-weighted opacities of the dominant neutral alkali metal lines at 1500 K and 1 bar.
This opacity spectrum has a bearing on the suggestion by BMS
that the strong continuum absorption seen
in all T dwarf spectra in the near-infrared from 0.8 \mic to 1.0 \mic, previously interpreted as due to
an anomalous population of red grains\cite{griffith} or in part due to
high-altitude silicate clouds\cite{allard}, is
most probably due to the strong red wings of the K I doublet at $\sim$7700 \AA.
This is demonstrated in Figure 2, in which several possible theoretical
spectra are compared with the observed spectrum for Gliese 229B in the near-infrared\cite{bms}.
Tsuji \etal\cite{tsuji99} also identified the K I doublet as one of the agents of absorption
shortward of one micron, but they needed silicate grains as well to reproduce the Gliese 229B observations.    
BMS conclude that the K I resonance doublet alone is responsible, though, given the
remaining ambiguity in its line shape, one can't completely eliminate
the presence of grains as secondary agents.

\begin{figure}[tf]
\begin{center}
\includegraphics[width=0.60\textwidth]{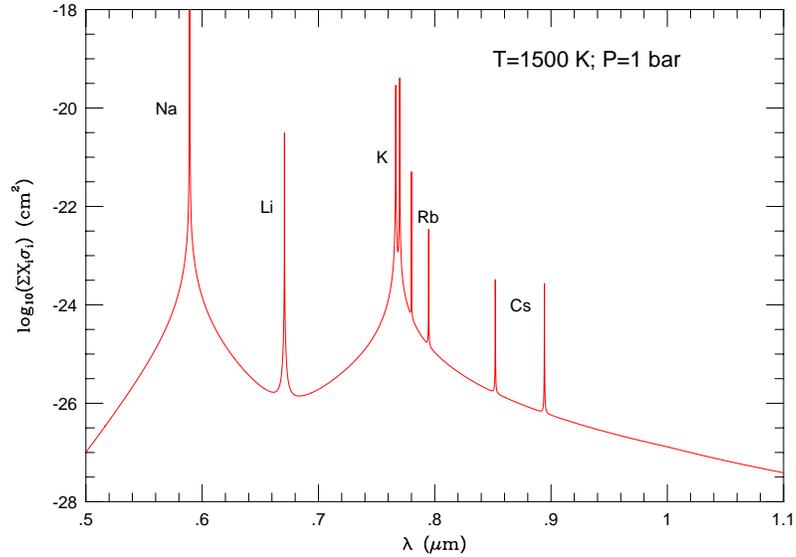}
\end{center}
\caption[]{Plotted is the abundance-weighted cross section spectrum for the
neutral alkali metals Na, K, Cs, Rb, and Li at 1500 K and 1 bar
pressure, using the theory of BMS.  The most important spectral lines for each species
are clearly marked.}
\label{fig1}
\end{figure}

\begin{figure}
\begin{center}
\includegraphics[width=0.8\textwidth]{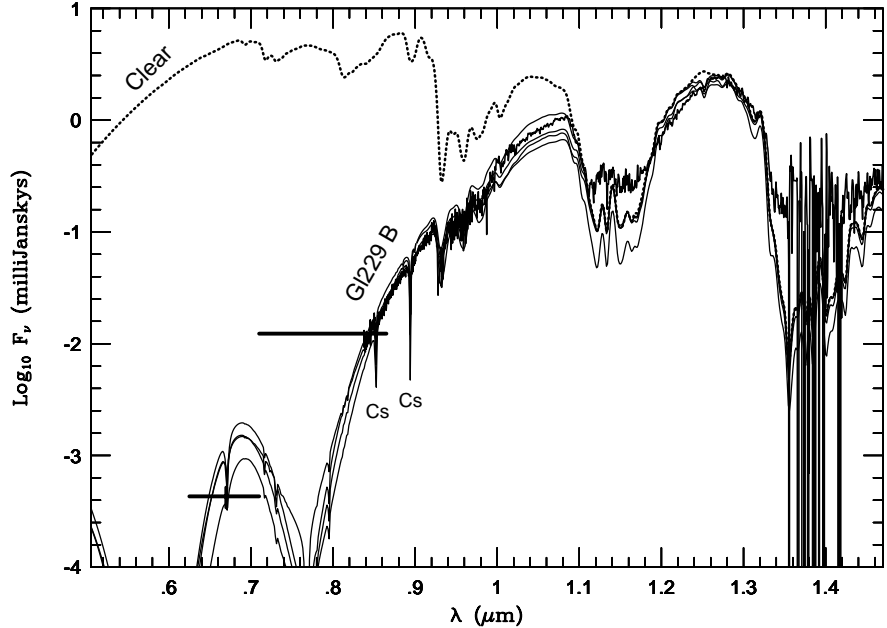}
\end{center}
\caption[]{The log of the absolute flux (F$_\nu$) in milliJanskys versus wavelength
($\lambda$) in microns from 0.5 \mic to 1.45 \mic for Gliese 229 B,
according to Leggett \etal \cite{leg99} (heavy solid), and for four theoretical models (light solid) described in BMS.  Also
included is a model, denoted ``Clear" (dotted),  without alkali metals and without any ad hoc absorber
due to grains or haze.  The horizontal bars near 0.7 \mic and 0.8 \mic denote the
WFPC2 $R$ and $I$ band measurements of Golimowski \etal\cite{golim}. Figure taken from BMS.}
\label{fig2}
\end{figure}

As Figure 2 suggests, the BMS theory also explains the WFPC2 $I$ band (M$_I\sim$20.76; theory = 21.0)
and $R$ band ($M_R \sim 24.0$; theory = 23.6) measurements made of Gl 229B\cite{golim},
with the Na D lines at 5890 \AA\ helping to determine the strength of the $R$ band.
BMS predicted not only that there would be a large trough in
a T dwarf spectrum at 7700 \AA\ due to the K I resonance, but that the spectrum of a T dwarf
would peak between the Na D and K I absorption troughs at 5890 \AA\ and 7700 \AA, respectively. This prediction
was recently verified by Liebert \etal\cite{liebert} for the T dwarf SDSS 1624+00.

Furthermore, the 1.17 \mic and 1.24 \mic subordinate lines of excited 
K I have been identified in T dwarfs\cite{strauss,tsvet,mclean}.
Since these subordinate lines are on the crown of
the $J$ band, they allow one to probe the deeper layers at higher temperatures.  Figure 3
portrays for a representative Gl 229B model the dependence on wavelength of the ``brightness''
temperature, here defined as the temperature at which the photon optical depth is $2/3$.
Such plots clearly reveal the temperature layers probed with spectra and
provide a means to qualitatively gauge composition profiles.
Specifically, for the Gl 229B model, the detection of the subordinate lines
of potassium indicates that we are there probing to $\sim$1600 K, while the detection of the fundamental methane
band at $3.3$ \mic (not shown in Figure 3) means that we are probing to only $\sim$600 K.

\begin{figure} 
\begin{center}
\includegraphics[width=0.60\textwidth]{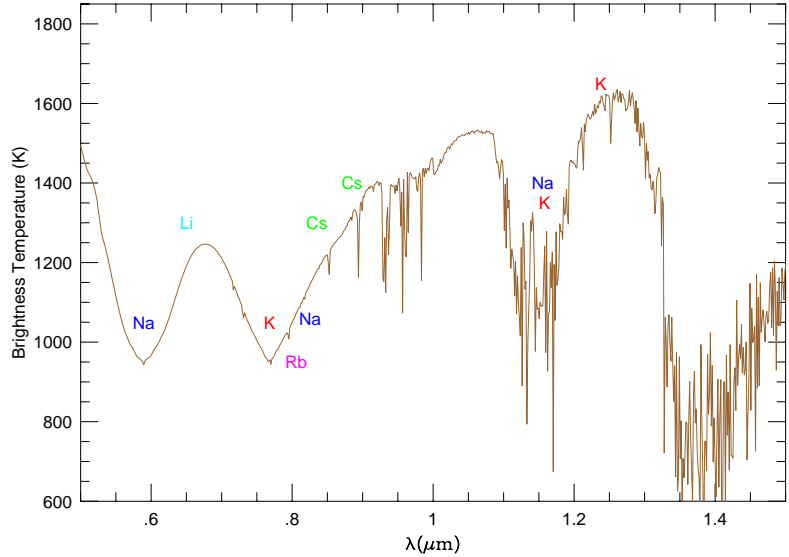}
\end{center}
\caption[]{The brightness temperature in Kelvin versus wavelength in microns
from 0.5 \mic to 1.5 \mic of a simple model of Gliese 229B.  The brightness 
temperature for a given wavelength is defined as the
temperature of the layer at which $\tau_{\lambda} = 2/3$.
The identity of the alkali metal atom responsible for a given feature is
indicated.  See text for discussion.
}
\label{fig3}
\end{figure}

\section{The Color of Brown Dwarfs}

Figure 1 shows that the Na D doublet should dominate the optical portion of the spectrum.
Since it suppresses the green wavelengths and ``brown'' is two parts red, one
part green, and very little blue, brown dwarfs should not be brown.  In fact, 
our recent calculations suggest that they are red to
purple, depending upon the exact shape of the line wings of Na D, the abundance
of the alkalis, the presence of high-altitude clouds, and the role of water clouds
at lower \teffs ($\sles$ 500 K).  A mixture of red and the complementary color 
to the yellow of the Na D line makes physical sense. It is the {\it complementary} color,
not the {\it color}, of the Na D line(s) because Na D is seen in absorption, not emission.
Indeed, the recent measurement of the spectrum of the
L5 dwarf 2MASSW J1507 from 0.4 \mic to 1.0 \mic (I.N. Reid and J.D. Kirkpatrick, in preparation)
indicates that this L dwarf is magenta in (optical) color.
%
%
This is easily shown with a program that generates the RGB equivalent of a given optical spectrum
(in this instance, R:G:B::1.0:0.3:0.42, depending upon the video ``gamma'').   
Hence, after a quarter century of speculation and ignorance,
we now have a handle on the true color of a brown dwarf --- and it is not brown.

\section{Rainout and the Alkali Metals}

As shown by Burrows and Sharp\cite{sharp}, Fegley and Lodders\cite{fegley}, and Lodders\cite{lodders}, the alkali metals
are less refractory than Ti, V, Ca, Si, Al, Fe, and Mg and survive in abundance as neutral atoms in substellar
atmospheres to temperatures of 1000 K to 1500 K.  This is below the 1600 K to 2500 K temperature
range in which the silicates, iron, the titanates, corundum, and spinel, etc. condense and rainout.
The rainout of refractory elements such as silicon and aluminum ensures that Na and K are not sequestered in the feldspars
high albite (NaAlSi$_3$O$_8$) and sanadine (KAlSi$_3$O$_8$) at temperatures
at and below 1400 K, but are in their elemental form down to $\sim$1000 K.
Hence, in the depleted atmospheres of the cool T dwarfs and late L dwarfs, alkali metals quite naturally
come into their own.  Figures 4 and 5 demonstrate the role of rainout by
depicting the profiles of the relative abundances of the main reservoirs
of the alkali metals, with and without rainout as crudely defined in reference \cite{sharp}.
As is clear from a comparison of these two figures, rainout and depletion of heavy metals can result
in a significant enhancement in the abundances at altitude (lower temperatures) of the neutral alkali metal atoms,
in particular sodium and potassium.

\begin{figure}
\begin{center}
\rotatebox{-90}{
\includegraphics[width=0.7\textwidth]{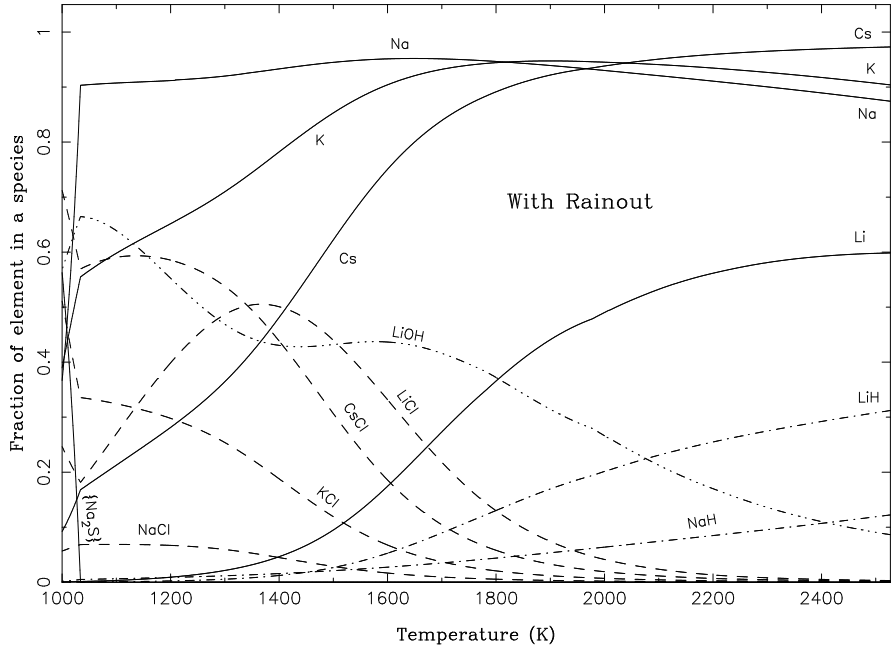}
}
\end{center}
\caption[]{The fractional abundances of different chemical species involving
the alkali elements Li, Na, K and Cs for a Gliese 229B model, with rainout
as described in Burrows and Sharp\cite{sharp}.
The temperature/pressure profile for a \teff=950 K and
$g=10^5$ cm s$^{-2}$ model, taken from Burrows \etal\cite{burr97}, was used.
Each curve shows the fraction of the alkali element in the indicated form
out of all species containing that element.
All species are in the gas phase except for the condensates, which are in braces \{
and \}.  The solid curves indicate the monatomic gaseous species Li, Na,
K and Cs, the dashed curves indicate the chlorides,
the dot-dashed curves indicate the hydrides and the triple dot-dashed curve indicates LiOH.
Due to rainout, at lower temperatures there is a dramatic difference
with the no--rainout, complete equilibrium calculation (Figure 5); high albite
and sanidine do not appear, but instead at a much lower temperature the
condensate Na$_2$S (disodium monosulfide) forms, as indicated by the solid
line in the lower left of the figure.  The potassium equivalent, K$_2$S, also
forms, but it does so below 1000 K and is not indicated here.
The difference between this Figure and Figure 5 is that almost all the silicon
and aluminum have been rained out at higher temperatures, so that no
high albite and sanidine form at lower temperatures. Figure taken from BMS.}
\label{fig4}
\end{figure}

\begin{figure}
\begin{center}
\rotatebox{-90}{
\includegraphics[width=0.7\textwidth]{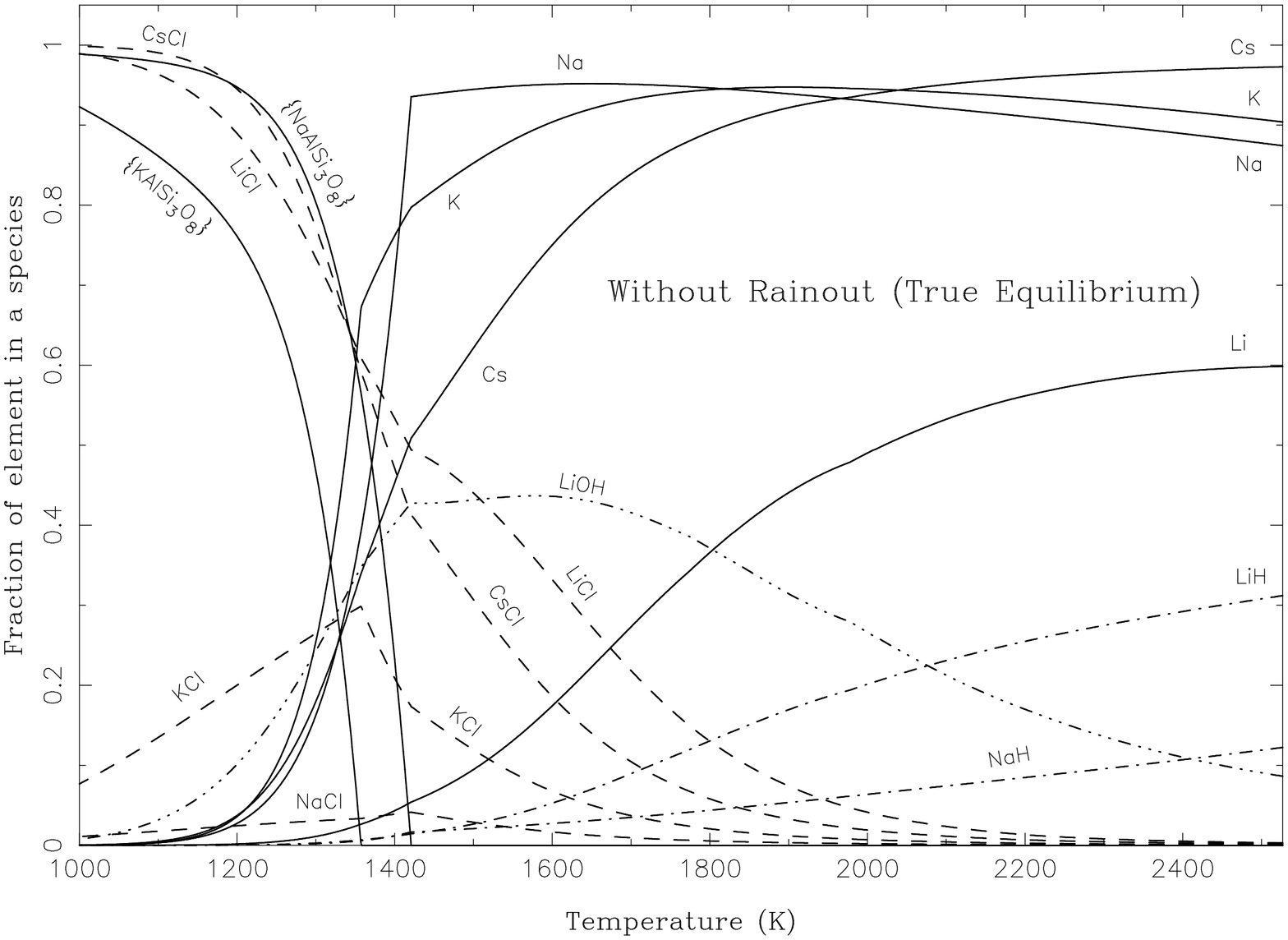}
}
\end{center}
\caption[]{The fractional abundances of different chemical species involving
the alkali elements Li, Na, K and Cs for a Gliese 229B model, assuming complete (true)
chemical equlibrium and no rainout (disfavored).  The temperature/pressure profile for a \teff=950 K and
$g=10^5$ cm s$^{-2}$ model, taken from Burrows \etal\cite{burr97}, was used.
Each curve shows the fraction of the alkali element in the indicated form
out of all species containing that element, {\it e.g.}, in the case of sodium, the
curves labeled as Na, NaCl, NaH and NaAlSi$_3$O$_8$ are the fractions of that
element in the form of the monatomic gas and three of its compounds.  All
species are in the gas phase except for the condensates, which are in braces \{
and \}.  The solid curves indicate the monatomic gaseous species Li, Na,
K and Cs and the two condensates NaAlSi$_3$O$_8$ and KAlSi$_3$O$_8$, {\it i.e.}, high
albite and sanidine, respectively, the dashed curves indicate the chlorides,
the dot-dashed curves indicate the hydrides and the triple dot-dashed curve
indicates LiOH. Figure taken from BMS.}
\label{fig5}
\end{figure}

Figure 2 demonstrates the naturalness
with which the potassium resonance lines alone fit
the observed near-infrared/optical spectrum of Gl 229B.
Curiously, in the metal-depleted atmospheres of T dwarfs the reach of the K I doublet
is one of the broadest in astrophysics, its far wings easily extending more
than 1500 \AA\ to the red and blue.  With rainout, below
$\sim$1000 K both sodium and potassium exist as sulfides (Na$_2$S and K$_2$S)\cite{lodders}.   Without rainout,
complete chemical equilibrium at low temperatures requires that sodium and potassium reside in
the feldspars.  If such compounds formed and persisted at altitude, then the
nascent alkali metals would be less visible, particularly in T dwarfs.  By modeling spectra with and
without the rainout of the refractories and comparing to the emerging 
library of T dwarf spectra\cite{mclean,burg99,burg00,leg00}, 
the degree of rainout and the alkali composition profiles 
in brown dwarf atmospheres can be approximately ascertained.

\section{Conclusion}

L and T dwarf spectra are unique among ``stars" and require new databases, approaches, and 
thinking to fully understand.  Exploring as we are new worlds, we will require new tools and
instincts with which to navigate.  Along with accurate cloud models, methane, and water, 
the alkali metals hold the key to unraveling the mysteries of the substellar objects that
we now know inhabit the solar neighborhood in abundance. 

\subsubsection{Acknowledgments:\\}
I thank my long-time collaborators, Jonathan Lunine, Bill Hubbard, and Mark Marley,
for simulating input and both Davy Kirkpatrick and Neill Reid for an advanced glimpse
at their stunning 2MASSW J1507 spectrum. This work was supported in part
by NASA under grants NAG5-7073 and NAG5-7499.


\begin{thebibliography}{17.}
\addcontentsline{toc}{section}{References}
 

\bibitem{leg99} S. Leggett, D.W. Toomey, T. Geballe, R.H. Brown: \apj \textbf{517}, L139 (1999)

\bibitem{bms} A. Burrows, M.S. Marley, C.M. Sharp: \apj \textbf{531}, 438 (2000).

\bibitem{griffith} C.A. Griffith, R.V. Yelle, M.S. Marley: Science \textbf{282}, 2063 (1998)

\bibitem{allard} F. Allard, P.H. Hauschildt, D.R. Alexander, S. Starrfield: \araa \textbf{35}, 137 (1997)

\bibitem{tsuji99} T. Tsuji, K. Ohnaka, W. Aoki: \apj \textbf{520}, L119 (1999)

\bibitem{golim} D.A. Golimowski, \etal: \aj \textbf{115}, 2579 (1998)

\bibitem{liebert} J. Liebert, I.N. Reid, A. Burrows, A.J. Burgasser, J.D. Kirkpatrick, J.E. Gizis: \apj \textbf{533}, 155 (2000)

\bibitem{strauss} M.A. Strauss, \etal: \apj \textbf{522}, L61 (1999)

\bibitem{tsvet} Z.I. Tsvetanov, \etal: \apj \textbf{531}, L61 (2000)

\bibitem{mclean} I.S. McLean, \etal: \apj \textbf{533}, L45 (2000)

\bibitem{sharp} A. Burrows, C.M. Sharp: \apj \textbf{512}, 843 (1999)

\bibitem{fegley} B. Fegley, K. Lodders: \apj \textbf{472}, L37 (1996)

\bibitem{lodders} K. Lodders: \apj \textbf{519}, 793 (1999)

\bibitem{burg99} A. Burgasser, \etal: \apj \textbf{522}, L65 (1999)

\bibitem{burg00} A. Burgasser, \etal: \apj \textbf{531}, L57 (2000)

\bibitem{leg00} S. Leggett, {\it et al.}: \apj \textbf{536}, L35 (2000)

\bibitem{burr97} A. Burrows, M. Marley, W.B. Hubbard, J.I. Lunine, T. Guillot, D. Saumon, R. Freedman,
D. Sudarsky, C. Sharp: \apj \textbf{491}, 856 (1997)

\end{thebibliography}
\end{document}